\documentclass[11pt]{article}
\usepackage{epsfig}
\begin{document}\setlength{\unitlength}{1mm}

\def\bef{\begin{figure}}
\def\eef{\end{figure}}
\newcommand{\ans}{ansatz }
\newcommand{\be}[1]{\begin{equation}\label{#1}}
\newcommand{\beq}{\begin{equation}}
\newcommand{\bea}{\begin{eqnarray}}
\newcommand{\ee}{\end{equation}}
\newcommand{\beqn}[1]{\begin{eqnarray}\label{#1}}
\newcommand{\eeqn}{\end{eqnarray}}
\newcommand{\bd}{\begin{displaymath}}
\newcommand{\ed}{\end{displaymath}}
\newcommand{\mat}[4]{\left(\begin{array}{cc}{#1}&{#2}\\{#3}&{#4}
\end{array}\right)}
\newcommand{\matr}[9]{\left(\begin{array}{ccc}{#1}&{#2}&{#3}\\
{#4}&{#5}&{#6}\\{#7}&{#8}&{#9}\end{array}\right)}
\newcommand{\matrr}[6]{\left(\begin{array}{cc}{#1}&{#2}\\
{#3}&{#4}\\{#5}&{#6}\end{array}\right)}
\newcommand{\cvb}[3]{#1^{#2}_{#3}}
\def\lsim{\raise0.3ex\hbox{$\;<$\kern-0.75em\raise-1.1ex
\hbox{$\sim\;$}}}
\def\gsim{\raise0.3ex\hbox{$\;>$\kern-0.75em\raise-1.1ex
\hbox{$\sim\;$}}}
\def\abs#1{\left| #1\right|}
\def\simlt{\mathrel{\lower2.5pt\vbox{\lineskip=0pt\baselineskip=0pt
           \hbox{$<$}\hbox{$\sim$}}}}
\def\simgt{\mathrel{\lower2.5pt\vbox{\lineskip=0pt\baselineskip=0pt
           \hbox{$>$}\hbox{$\sim$}}}}
\def\unity{{\hbox{1\kern-.8mm l}}}
\def\epr{E^\prime}
\newcommand{\al}{\alpha}
\def\16p{16\pi^2}
\newcommand{\eps}{\varepsilon}
\newcommand{\epsr}{\varepsilon_{ R}}
\newcommand{\epsl}{\varepsilon_{ L}}
\newcommand{\epsrs}{\varepsilon_{s R}}
\newcommand{\epsls}{\varepsilon_{s L}}
\def\ep{\epsilon}
\def\ga{\gamma}
\def\Ga{\Gamma}
\def\om{\omega}
\def\OM{\Omega}
\def\la{\lambda}
\def\La{\Lambda}
\def\al{\alpha}
\newcommand{\ov}{\overline}
\renewcommand{\to}{\rightarrow}
\renewcommand{\vec}[1]{\mbox{\boldmath$#1$}}
\def\tm{{\widetilde{m}}}
\def\mcirc{{\stackrel{o}{m}}}
\def\dem{\delta m^2} 
\def\sint{\sin^2 2\theta} 
\def\tant{\tan 2\theta} 
\def\tanL{\tan 2\theta^L}
\def\tanR{\tan 2\theta^R}
\newcommand{\tanb}{\tan\beta}
\def\brf{{\mathbf f}}
\def\bbf{\bar{\bf f}}
\def\bF{{\bf F}}
\def\bbF{\bar{\bf F}}
\def\bFp{{\bf F^\prime}}
\def\bbFp{\bar{\bf F^\prime}}
\def\bY{{\mathbf Y}}
\def\by{{\mathbf y}}
\def\bX{{\mathbf X}}
\def\bS{{\mathbf S}}
\def\bM{{\mathbf M}}
\def\bA{{\mathbf A}}
\def\bB{{\mathbf B}}
\def\bG{{\mathbf G}}
\def\bI{{\mathbf I}}
\def\bb{{\mathbf b}}
\def\bh{{\mathbf h}}
\def\bg{{\mathbf g}}
\def\bla{{\mathbf \la}}
\def\bmu{\mathbf m }
\def\bunity{{\mathbf 1}}
\def\cA{{\cal A}}
\def\cB{{\cal B}}
\def\cC{{\cal C}}
\def\cD{{\cal D}}
\def\cF{{\cal F}}
\def\cG{{\cal G}}
\def\cH{{\cal H}}
\def\cI{{\cal I}}
\def\cL{{\cal L}}
\def\cO{{\cal O}}
\def\cR{{\cal R}}
\def\cS{{\cal S}}
\def\cT{{\cal T}}
\def\dfrac#1#2{{\displaystyle\frac{#1}{#2}}}
\newcommand{\tphi}{\tilde{\phi}}
\newcommand{\Tphi}{\tilde{\Phi}}
\def\Bsi{{\bar{\Psi}}}
\newcommand{\bx}{\bar{\rm X}} 
\newcommand{\wx}{{\rm X}} 
\newcommand{\bv}{\bar{\rm V}} 
\newcommand{\wv}{{\rm V}} 
\newcommand{\tl}{\tilde{l}} 
\newcommand{\tq}{\tilde{q}}
\newcommand{\tu}{\tilde{u}}
\newcommand{\td}{\tilde{d}}
\newcommand{\tuc}{\tilde{u}_c} 
\newcommand{\tdc}{\tilde{d}_c} 
\newcommand{\tec}{\tilde{e}_c} 
\newcommand{\TQ}{\tilde{Q}} 
\newcommand{\TU}{\tilde{U}}
\newcommand{\TE}{\tilde{E}} 
\newcommand{\TUC}{\tilde{U}_c} 
\newcommand{\TEC}{\tilde{E}_c} 
\newcommand{\TQC}{\tilde{Q}_c} 
%\renewcommand{\thefootnote}{\fnsymbol{footnote}}
%\def\thefootnote{\fnsymbol{footnote}}
%
%%%%%%%%%%%% end my definitions %%%%%%%%%%%%%%%%%%%%%%%%%%%%%%%%%

\begin{titlepage}

\begin{flushright}
DFAQ-01/TH/08 \\ 
DFPD-01/TH/46 \\ 

%\today
\end{flushright}

\vspace{2.0cm}

\begin{center}

{\Large \bf
Limits on the Non-Standard  Interactions of Neutrinos   \\
\vspace{0.4cm}
from $e^+ e^-$ Colliders}

\vspace{0.7cm}

{\large \bf Zurab Berezhiani${}^{a,b,}$\footnote{
E-mail address: berezhiani@fe.infn.it } 
and Anna Rossi${}^{c,}$\footnote{
E-mail address: arossi@pd.infn.it } 
}
\vspace{5mm}

{\it ${}^a$ Dipartimento di Fisica, 
Universit\`a di L'Aquila,  
I-67010 Coppito, AQ, and \\
INFN, Laboratori Nazionali del Gran Sasso, I-67010 Assergi, AQ, Italy}

{\it ${}^b$ The Andronikashvili Institute of Physics, 
Georgian Academy of Sciences, \\ 
380077 Tbilisi, Georgia}

{\it  ${}^c$ Dipartimento di Fisica, 
Universit\`a di Padova and 
INFN Sezione di Padova, \\ I-35131 Padova, Italy. 
}
\end{center}

\vspace{10mm}

\begin{abstract}
\noindent
We provide an effective Lagrangian analysis 
of contact non-standard interactions 
of neutrinos with electrons, which can be effectively mediated 
by extra particles, and 
 examine the associated experimental limits. 
At present,  such 
interactions are strongly constrained only for $\nu_\mu$: 
the bounds are loose for $\nu_e$ and absent for $\nu_\tau$.
We emphasize the unique role played by the reaction 
$e^+e^-\rightarrow \nu \bar{\nu}\gamma$  in providing 
direct constraints on such non-standard  interactions.

\end{abstract}
\end{titlepage}
\setcounter{footnote}{0}

%%%%%%%%%begin

\section{Introduction}
Nowadays neutrino physics is very alive thank to 
the exciting measurements and results reported by atmospheric and 
solar neutrino experiments.
They have been confirming that  
neutrinos do oscillate and hence they may be   mixed and massive, 
offering us in this way  important informations on the `flavour' 
structure of the Standard Model (SM). 
In general, extensions of the SM are required to generate non-vanishing 
neutrino masses and mixing angles, which are supposed to originate 
at an energy scale much larger than the electroweak scale, for instance
the grand unified scale ( $\sim 10^{16}$ GeV).
Then, as neutrino physics is susceptible to be extended beyond the 
SM realm, 
it is also conceivable that some 
new physics may also predict 
novel interactions of neutrinos with matter constituents which 
can be flavour changing  as well as flavour conserving. 
The phenomenological 
relevance of such non-standard (NS) interactions will depend in general 
on the scale at which they are presumed to arise: it should be not 
too far from the electroweak scale.   
Non-standard flavour changing neutrino interactions 
have been invoked long ago to explain the solar neutrino anomaly 
(SNA) \cite{FC,AZ1,AZ2,analysis}. 
Some time ago  \cite{AZ1}   the impact of extra 
$\nu_\tau$ interactions with electrons on the detection cross section  
 has been also investigated 
in the context of the neutrino long wavelength 
oscillation as a solution to the SNA.
The interest in neutrino NS interactions 
continues to increase.\footnote{We have to mention that 
very recently the 
results by the NuTeV experiment on the determination 
of electroweak parameters 
show a discrepancy with the SM expectation 
that suggests non-standard 
couplings of neutrinos with quarks \cite{nutev}.
}
In particular, at present  two 
points of view can be taken: either to use solar or atmospheric neutrino 
data to constrain neutrino NS interactions \cite{analysis,fornengo}, or, on the contrary, 
to use solar  neutrino experiments to detect signatures of neutrino NS 
interactions \cite{BRR}. 
Both approaches cannot leave aside the constraints that 
emerge from laboratory experiments \cite{BPW}.  
The properties of $\nu_e$ and $\nu_\mu$ have been tested and  while 
the accelerator constraints on $\nu_e$ are still loose, those on $\nu_\mu$ 
are rather severe \cite{BPW}.  
On the contrary, until now  the $\nu_\tau$ properties have not 
been directly   tested 
in laboratory. Indeed, the constraints on non-standard interactions 
rather apply to its $SU(2)_W$-partner, the $\tau$ lepton.
In this paper, we point out that neutrino NS interactions with electrons 
can be constrained at $e^+ e^-$ colliders through the reaction  
$e^+ e^- \to \nu\bar{\nu}\ga$.\footnote{Limits on the magnetic moment of 
the $\tau$ neutrino have been provided using this reaction \cite{sarma}.} 
It is well-known that the 
invisible channel reaction with photon emission provides 
a useful tool for measuring the number of light 
neutrinos \cite{DOZ} and for revealing new physics. 
We note that the constraints 
derived from the cross section measurements performed at LEP are
the only laboratory bounds on the $\nu_\tau$ NS neutral current interactions 
and may be competitive to those obtained from $\nu_e e$ scattering 
as regards the $\nu_e$ NS interactions. 
(The same bounds when applied to $\nu_\mu$ are not competitive 
with the existing ones from the $\nu_{\mu}  e$ elastic scattering 
 \cite{charm2}.)

We shall find that $\nu_\tau$ may have sizeable extra-interactions 
with electrons. 
This has recently pushed us \cite{BRR} to investigate in detail the 
 possibility  to `identify' the $\nu_\tau$ in solar 
neutrino elastic scattering experiments like Borexino \cite{BX}. 
Indeed, in view of the atmospheric neutrino data 
pointing to a quasi-maximal mixing between $\nu_\mu$ and $\nu_\tau$, 
 the solar neutrino anomaly is to be interpreted as the 
conversion of  $\nu_e$ into a  state of nearly equal mixture of 
$\nu_\mu$ and $\nu_\tau$. Therefore the Sun is an abundant source of 
$\tau$ neutrinos. While in the framework of 
the Standard Model this observation may sound `academic', it becomes 
meaningful as soon as $\nu_\tau$ is allowed to 
have extra (neutral current) 
interactions and then  to be distinguishable  from $\nu_\mu$. 
The elastic  scattering $\nu e$ that will be used by Borexino to detect 
the mono-energetic $^7Be$  neutrinos can be in this case the right 
place where to observe neutrino NS interactions with electrons. 
They could show up in some specific deformations of the electron energy spectrum. 
The observation of such an effect 
% distorted energy spectrum 
would simultaneously 
provide the signature of $\nu_\tau$ interactions and would be a 
further proof of 
the large mixing angle i.e. a test of  
the atmospheric neutrino oscillations. 

The content of this paper is organized as follows. 
In Sec. 2 we present the effective Lagrangian describing 
neutrino NS interactions with electrons.
Sec. 2.1  provides 
a derivation of the effective $SU(2)_W \times U(1)_Y$ invariant operators 
(with $d\geq 6$) 
describing neutrino interactions with electrons. 
This enables us to find possible theoretical realizations 
of such interactions without conflicting with the experimental 
bounds (Sec. 2.2). In Sec. 3 we briefly review the present  
laboratory bounds  on  NS leptonic interactions.
So we focus 
on the bounds on $\nu$-$e$ NS interactions achievable
with the measurements of the cross section 
$e^+e^-\to \nu \bar{\nu}\gamma$ (Sec. 4).
Our findings are summarised in Sec. 5.

\section{Non-standard leptonic interactions}

Our discussion is most focused on neutrino interactions with electrons.
In the Standard Model, the elastic scattering $\nu_\al e\to\nu_\al e$
of neutrinos with the electron  are  described at low energies by the
following four-fermion operator ($\nu_\al = \nu_e, \nu_\mu, \nu_\tau$):
\be{SM}
-{\cal L}^\nu_{\rm SM} = 
{2\sqrt2 G_F} (\ov{\nu}_\al \gamma^\mu P_L \nu_\al)
\left[\, g_R\ov{e}\gamma_\mu P_R e +
g_L\ov{e}\gamma_\mu P_L e   \, \right]
\, + \,
{2\sqrt2 G_F} (\ov{\nu}_e \gamma^\mu P_L \nu_e)  
(\ov{e}  \gamma_\mu P_L e)
\ee
where $G_F$ is the Fermi constant, $P_{L,R}= (1\mp \ga^5)/2$
are the chiral projectors, $g_R = \sin^2\theta_W$ and
$g_L= -\frac12 +\sin^2\theta_W$
are the electron coupling constant to the $Z$-boson,
and the last term is the additional contribution
for the electron neutrino arising from  the $W$-boson exchange.
Alternatively, we can define the vector and axial coupling constant 
$g_{V} = g_{L} + g_{R}$ and $g_{A} = g_{L} - g_{R}$ and properly 
express the corresponding four-fermion interaction in terms of them. 

We assume on phenomenological grounds that neutrinos may also have
 NS interactions with the electron described
by the following four-fermion operator:\footnote{For simplicity, 
here we consider only  neutrino flavour-diagonal interactions 
with electrons, though in general there could also exist 
 the  flavour-changing ones.}  
\be{NS}
-{\cal L}^\nu_{\rm NS} =
{2\sqrt2G_F} (\ov{\nu}_\al \gamma^\mu P_L \nu_\al)
\left[\, \eps_{\al R}  \ov{e} \gamma_\mu P_R e
+ \eps_{\al L} \ov{e}  \gamma_\mu P_L e \, \right] ,   
\ee
where $\eps_{\al R}$ and $\eps_{\al L}$ parameterize the strength   
of the new interactions with respect to $G_F$, and
in general they can be dependent   
on the neutrino flavour ($\al =e,\mu,\tau$).
The occurrence of these  extra interactions entails
a `coherent' redefinition of the constants $g_{L,R}$
\be{NCnutau}
g_{R} \to \tilde{g}_{\al R} = g_{R} + \eps_{\al R} , ~~~~~~~
g_{L} \to \tilde{g}_{\al L} = g_{L} + \eps_{\al L}.
\ee
so that the effective coupling constants 
$\tilde{g}_{\al R}$ and $\tilde{g}_{\al L}$
in general depend on the flavour of the neutrino. 
Equivalently, we can define the vector and axial constants
of the extra interactions, 
$\eps_{\al V} = \eps_{\al L} + \eps_{\al R}$ and 
$\eps_{\al A} = \eps_{\al L} - \eps_{\al R}$.  
Although this presentation may appear redundant, in the next    
we shall explicitly display the allowed parameter space for both
($\eps_{R},\eps_{L}$) and ($\eps_{V},\eps_{A}$)
while discussing the bounds on the non-standard neutrino couplings.
Indeed, the former parametrization is closer to the theoretical
perception, while the latter is somehow more appropriate
to the neutrino-propagation phenomenology in matter.

\subsection{Theoretical aspects: effective operator picture} 
In the framework of the Standard Model,  neutrino NS 
interactions (\ref{NS}) with electrons can be derived by  
$SU(2)_W\times U(1)_Y$-invariant operators  
of dimension $d\geq 6$:
\be{serie}
\dfrac{h^{\al(n)}_{R}(H,H^\dagger) }{M^2} 
[ (\bar{l}_\al\ga^\mu P_L l_\al)
(\bar{e}\ga_\mu P_R e) ]_{(n)} +
%(\ov{l}_\tau \ov{e}^c)( e^c l_\tau) +
\dfrac{h^{\al(n)}_{L}(H,H^\dagger) }{M^2} 
[ (\bar{l}_\al\ga^\mu P_L l_\al)
(\bar{l}_e\ga_\mu P_L l_e) ]_{(n)} , 
\ee
where
$l_{\al}$ denotes the  the $SU(2)_W$ 
lepton doublets ($\al =e, \mu, \tau$),
$H$ is the Higgs doublet and 
$M$ is a cutoff mass scale.\footnote{For an earlier 
general analysis of $d = 6$ effective operators, 
see e.g. \cite{BW}.} 
In the above expressions all possible $SU(2)_W$ contractions 
are implicitly understood and counted by the index $n$, 
while the round brackets merely indicate  Lorentz contractions. 
The field-dependent couplings $h^{\al(n)}_{R,L}$ 
are to be understood as follows ($A=R,L$):
\be{ghh}
h^{\al(n)}_{A}(H,H^\dagger)= h^{\al(n)}_{0, A}+ h^{\al(n)}_{1,A}
\dfrac{(H H^\dagger)_{n}}{M^2} + \cdots ~.
\ee
The $h^{\al(n)}_{R}$-expansion
will give neutrino interactions with right-handed 
electrons ($\eps_{\al R}$-coupling), while the $h^{\al(n)}_{L}$-expansion  
will give neutrino interactions with left-handed 
electrons ($\eps_{\al L}$-coupling).
In eq.~(\ref{ghh}) the dots  stand for higher-dimension operators 
that we omit for brevity since 
the explicit discussion of $d=6$ and $d=8$ operators is already sufficient 
to illustrate the main points.
We can easily see that the $h^{\al(n)}_R$-type expansion (\ref{ghh})  
produces one invariant ($n=1$)
at lowest order in $1/M^2$, 
and two invariants ($n=1,2$) at next order:
\beqn{Rinv}
&& h^{\al(1)}_{0,R} (l^\dagger_\al l_\al) (e^\dagger e) , \nonumber \\
&& h^{\al(1)}_{1,R}S (l^\dagger_\al l_\al)(e^\dagger e) +
%(H^\dagger H)(\cdots) +
h^{\al(2)}_{1,R} T^a (l^\dagger_\al \sigma^a l_\al)(e^\dagger e) .
\eeqn
Here the Lorentz structure is understood and we have defined 
$S \equiv (H^\dagger  H)/M^2~$,  
$~T^a \equiv (H^\dagger \sigma^a H)/M^2$ 
($\sigma^a$ are the Pauli matrices, $a=1,2,3$).
On the other hand, the  $h^{\al(n)}_L$-type expansion  entails 
more invariants, namely ($\al =e,\mu,\tau$; $\beta =\mu,\tau$): 
\bea
&& h^{\al(1)}_{0,L} (l^\dagger_\al l_\al)(l^\dagger_e l_e)  , 
  \nonumber \\
&& h^{\beta(2)}_{0,L} (l^\dagger_\beta \sigma^a l_\beta)
(l^\dagger_e \sigma^a l_e) , 
\eeqn
\be{L1inv}
\begin{array}{lllll}
 h^{\beta(1)}_{1,L}S (l^\dagger_\beta l_\beta)
(l^\dagger_e l_e)  , ~~~~~&
 h^{e(1)}_{1,L}S (l^\dagger_e l_e)(l^\dagger_e l_e)  , 
\nonumber \\
h^{\beta(2)}_{1,L}S 
(l^\dagger_\beta \sigma^a l_\beta)(l^\dagger_e \sigma^a l_e)  , ~~~~~&
 h^{e(2)}_{1,L}T^a (l^\dagger_e \sigma^a l_e)
(l^\dagger_e l_e) ,
\nonumber \\
 h^{\beta(3)}_{1,L} T^a (l^\dagger_\beta \sigma^a l_\beta)
(l^\dagger_e l_e)  ,
\nonumber \\
h^{\beta(4)}_{1,L}T^a (l^\dagger_\beta  l_\beta)
(l^\dagger_e \sigma^a l_e)  ,~~~~~&
\nonumber \\
h^{\beta(5)}_{1,L}\eps_{abc} T^a (l^\dagger_\beta \sigma^b l_\beta)
(l^\dagger_e \sigma^c l_e)  .~~~~~& 
\end{array} ~ 
\ee
The operators (\ref{Rinv}-\ref{L1inv}) form a particular 
basis in which the singlet and triplet $SU(2)_W$-orientation are 
manifest, as the symbols $S$ and $T^a$ adopted for the 
dimensionless Higgs  combination  suggest,    
but there are of course other equivalent bases linearly related 
to this one.\footnote{We mention that the invariant formed by the  
$SU(2)_W$-quintuplet  orientation, i.e $1 \subset 5 \times 5$ appears at 
the successive order, 
$d=10$. The  $SU(2)_W$-quintuplet representation is the highest one 
that can be realized, throughout all the expansion. 
However, this $SU(2)_W$ structure does not give 
rise to further effective interactions 
in the broken phase (see below, eq.~\ref{NS-2}), besides those 
derived from the $SU(2)_W$  
singlet and triplet operator structure 
in eqs.~(\ref{Rinv}-\ref{L1inv}).}
Once $SU(2)_W\times U(1)_Y$ is broken by 
the vacuum expectation value (VEV)  of 
the Higgs field $H$,  $\langle H^0\rangle =v$, 
 all the operators above 
become $d=6$ four-fermion operators of the broken phase. 
These  effective operators  
can be cast into a parameterization like that  in eq.~(\ref{NS}), namely
\beqn{NS-2}
- {\cal L}^\nu_{eff}& =& 
{2\sqrt2G_F} (\bar{\nu}_\al \gamma^\mu P_L \nu_\al)
\left[ \eps_{\al R} (\bar{e}  \gamma_\mu P_R e) + 
\eps_{\al L}  (\bar{e}  \gamma_\mu P_L e) \right] , 
\nonumber \\    
- {\cal L}^{(1)}_{eff} & = & 
{2\sqrt2G_F} (\bar{\tau} \gamma^\mu P_L \tau )
\left[ \kappa_{\tau R} (\bar{e}  \gamma_\mu P_R e) + 
\kappa_{\tau L}  (\bar{e}  \gamma_\mu P_L e) \right] + 
(\tau \to \mu) + (\tau \to e) , 
\nonumber \\
- {\cal L}^{(2)}_{eff} & = & 
{2\sqrt2G_F} (\bar{\nu}_\tau \gamma^\mu P_L \nu_e)
\left[\zeta_{\tau L}(\bar{e}\gamma_\mu P_L \tau) 
%+ \zeta_{\tau R}(\bar{e}\gamma_\mu P_R \tau) 
+ \mbox{h.c.} \right]
+(\tau\to \mu) , 
\nonumber \\
- {\cal L}^{(3)}_{eff} & = & 
{2\sqrt2G_F} (\bar{\nu}_e  \gamma_\mu P_L \nu_e) 
\left[ \xi_{\tau L} (\bar{\tau} \gamma^\mu P_L \tau) 
+\xi_{\nu_\tau L} (\bar{\nu}_{\tau} \gamma^\mu P_L \nu_\tau) \right]
   +(\tau \to \mu)   
\nonumber \\
&& + {2\sqrt2G_F}\xi_{\nu_e L} (\bar{\nu}_e  \gamma_\mu P_L \nu_e)  
(\bar{\nu}_e  \gamma_\mu P_L \nu_e )  , 
\eeqn
where the dimensionless parameters such as 
$\eps_{\al R (L)}$,  $\kappa_{\al R (L)}$ etc. 
are to be identified  as follows ($\al=e,\mu,\tau ; \beta=\mu,\tau$):
\beqn{eps}
2\sqrt2G_F \eps_{\al R}&= & \frac{1}{M^2}\left[
{h^{\al (1)}_{0,R}} +  S h^{\al (1)}_{1,R} +T h^{\al  (2)}_{1,R}  
  +\cdots \right] , 
\nonumber \\
2\sqrt2G_F \kappa_{\al R}&= &\frac{1}{M^2}\left[ 
{h^{\al (1)}_{0,R}} + 
 S h^{\al (1)}_{1,R} - T h^{\al (2)}_{1,R} 
 +\cdots \right], \nonumber \\
 2\sqrt2G_F \eps_{\beta  L}& = & \frac{1}{M^2}\left[
{(h^{\beta (1)}_{0,L} - h^{\beta (2)}_{0,L} ) } + 
 S(h^{\beta (1)}_{1,L} - h^{\beta (2)}_{1,L})+  
T(h^{\beta (3)}_{1,L} - h^{\beta (4)}_{1,L})
 + \cdots \right] ,  \nonumber \\
2\sqrt2G_F \eps_{e L}& =  &\dfrac{2}{M^2}\left[
h^{e(1)}_{0,L}  + S h^{e (1)}_{1,L} 
 + \cdots \right] , \nonumber \\ 
2\sqrt2G_F \kappa_{\beta L}&= & \frac{1}{M^2}\left[ 
{(h^{\beta (1)}_{0,L} + h^{\beta (2)}_{0,L} ) } + 
 S(h^{\beta (1)}_{1,L} + h^{\beta (2)}_{1,L}) - T  
(h^{\beta (3)}_{1,L} + h^{\beta (4)}_{1,L})
 + \cdots \right] ,  \nonumber \\
2\sqrt2G_F \kappa_{e L}&= &  \frac{1}{M^2}\left[
h^{e(1)}_{0,L}  + 
S h^{e (1)}_{1,L} - T h^{e(2)}_{1,L} 
 + \cdots \right]  , \nonumber \\
2\sqrt2G_F \zeta_{\beta L} & = &\frac{2}{M^2}\left[
{h^{\beta (2)}_{0,L} } +
S h^{\beta(2)}_{1,L} + i T h^{\beta(5)}_{1,L} 
  + \cdots \right]  , \nonumber \\
 {2\sqrt2G_F}  \xi_{\beta L} &= & \frac{1}{M^2}\left[
{(h^{\beta(1)}_{0,L} - h^{\beta(2)}_{0,L} ) }
+ 
S (h^{\beta(1)}_{1,L} - h^{\beta(2)}_{1,L} )-T  
(h^{\beta(3)}_{1,L} - h^{\beta(4)}_{1,L})
 + \cdots \right] \nonumber \\
{2\sqrt2G_F}  \xi_{\nu_{\beta } L} &=  
&\frac{1}{M^2}\left[
{(h^{\beta(1)}_{0,L} + h^{\beta(2)}_{0,L} ) }
+ 
S(h^{\beta(1)}_{1,L} + h^{\beta(2)}_{1,L}) + T  
(h^{\beta(3)}_{1,L} + h^{\beta(4)}_{1,L})
 + \cdots \right] \nonumber \\
%% for nue
2\sqrt2G_F \xi_{\nu_e L}&= & \frac{1}{M^2}\left[
h^{e(1)}_{0,L} +
S h^{e (1)}_{1,L} + T h^{e(2)}_{1,L} 
   + \cdots \right] 
\eeqn
It is understood that now the symbols $S$ and $T$ 
stand for $\langle S\rangle = v^2/M^2$ and  
$\langle T^3\rangle= -2 Y_H v^2/M^2$ 
(where $Y_H$ is the $H$ hypercharge), respectively. 
Notice that in the triplet-orientation $T^a$ only the third component 
contributes.
Also notice that the terms proportional to $S$ could be reabsorbed 
in the lowest-order couplings $h^{\al (n)}_{0,A}$.

We have collected  in  ${\cal L}^{\nu}_{eff}$ 
all the desired operators 
involving only the neutrino current with either 
the right-handed or the left-handed electron current, eq.~(\ref{NS}). 
However, we have more interactions which involve  also the other 
components of the $SU(2)_W$ doublets.   
Among these we distinguish  those which are  `observable',   
contained  in the Lagrangians ${\cal L}^{(1)}_{eff}$,  
${\cal L}^{(2)}_{eff}$, and all other interactions  which instead are 
`unobservable', collected in ${\cal L}^{(3)}_{eff}$.
Here by `observable' we mean operators which can give phenomenologically 
testable interactions and then are subjected to constraints.\footnote{
However,  a process such as  
$\mu^+ \mu^- \to \nu_e \bar{\nu}_e$ described by ${\cal L}^{(3)}_{eff}$ 
could be studied at the planned muon colliders. 
As for the  $\xi_{\nu_\al}$-like neutrino interactions 
in  ${\cal L}^{(3)}_{eff}$,
they can contribute to the decay $Z\to \nu\bar{\nu}$ at one-loop and 
thus can be constrained by 
the invisible $Z$-width determination, $\xi_{\nu_\al}\lsim 1$  \cite{BS}. 
Analogous considerations 
for the decays $Z\to \ell \bar{\ell}$ constrain the couplings $\eps_{R,L}, 
\kappa_{R,L}$ as $\eps_{R,L}, 
\kappa_{R,L} < 0.5$ or so. These bounds are only meant as estimates, 
since such loop effects are cut-off dependent.} 

Some interesting features emerge. 
First, consider the limit of unbroken $SU(2)_W$. Then only
the $d=6$ operators are relevant. The two types of
interactions with right-handed electrons have the
same coupling constants ($\eps_{\al R}=\kappa_{\al R}$).
As for the interactions with left-handed electrons,
for $\al = e$ we have three interactions with 
the same coupling constant,
whereas for $\beta = \mu,\tau$ we have five interactions
determined by only two coupling constants.
In this limit, therefore, 
the stringent laboratory bounds on the  
effective couplings $\kappa_{\al L}, \kappa_{\al R}$ and 
$\zeta_{\beta L}$ in  ${\cal L}^{(1)}_{eff}$ and ${\cal L}^{(2)}_{eff}$
would imply similar bounds also for the couplings   
$\eps_{\al L}, \eps_{\al R}$ in ${\cal L}^\nu_{eff}$,
which by themselves are more loosely constrained. 
(For a recent discussion on this issue see \cite{BGP}.)
Now consider the effects induced by the breaking
of $SU(2)_W$. Then also the $d > 6$ operators play a role
and can correct (or even spoil) the strong correlations
among couplings described above\footnote{
Needless to say that the emergence of {\it large} 
$SU(2)_W$-breaking effects means that in the expansion (\ref{serie}) 
all higher order terms can become comparable in magnitude to the leading 
one, in which case a proper re-summation has to be implemented. 
Alternatively, 
once the fundamental theory is given, one could include $SU(2)_W$-breaking 
effects in an exact, rather than approximate, way.}.   
These features are manifest in eq.~(\ref{eps}).
It is then conceivable that the effective couplings
in ${\cal L}^{(1)}_{eff}$ and ${\cal L}^{(2)}_{eff}$ 
be suppressed below the experimental bounds, while
at the same time the couplings in ${\cal L}^\nu_{eff}$ 
remain sizeable. Notice also that for example 
a sizeable $\eps_{e L}$ could be compatible with small $\kappa_{e L}, 
\zeta_{eL}$ but would also imply that the  four-neutrino 
interaction parameter $\xi_{\nu_e}$ be  large. This may have
interesting consequences for the neutrino decoupling 
in the Early Universe as well as for the 
neutrino propagation in dense matter, like that in the 
supernova core.
In conclusion, it is reasonable and worthwhile to explore 
the possible experimental relevance of neutrino 
non-standard interactions \cite{BRR}.
 
In the following we shall disregard the second-generation 
($\nu_\mu$, $\mu$)-couplings as these 
are already severely bounded,  
$\eps_{\mu R,L} <0.1$ and $\kappa_{\mu R,L}, \zeta_{\mu L} < 0.01$.
So we shall focus mostly on the extra interactions of $\nu_e$ and 
$\nu_\tau$ for 
which the parameters $\eps_{e R}, \eps_{e L}$ and  
$\eps_{\tau R}, \eps_{\tau L}$ 
are so far weakly bounded or unbounded, respectively.

\subsection{Explicit examples for neutrino non-standard interactions}
 
We shall briefly give two examples of interactions in the 
fundamental theory, at energy-scale larger than $\Lambda$,   
to show how the $SU(2)_W$-universality can be concretely broken, 
specializing, 
for the sake of simplicity, to the third generation case.  

\noindent
$\bullet$ By exchanging an additional scalar $SU(2)_W$ doublet $\phi$  
 according to the following coupling:
\be{phi1}
\bar{l}_\tau P_R e \phi +{\mathrm h.c.}
\ee
we generate the leading $d=6$ operators   and so 
the first term in the expressions of  $\eps_{\tau R}$  and 
$\kappa_{\tau R}$ (\ref{NS-2}) in  ${\cal L}^\nu_{eff}$ and 
${\cal L}^{(1)}_{eff}$, respectively. 
If $H$-legs are inserted  in the $\phi$-propagator,  
higher-dimension operators ($d=8$ etc.)  are effectively generated. 
The non-vanishing VEV of $H$ introduces $SU(2)_W$ breaking effects, 
as already outlined, and so generates the successive terms in the 
$\eps_{\tau R}$ and $\kappa_{\tau R}$-series.  
This amounts to saying that $SU(2)_W$ breaking effects split the 
masses of the components of the $\phi$ doublet,  
$M_{\phi^+}\neq M_{\phi^0}$, and thus generate 
different effective couplings for the interactions  of $\nu_\tau$ and its 
partner, the $\tau$ lepton, with the electron.
However, this mass splitting would contribute to the 
Standard Model $\rho$ parameter and the maximally allowed 
ratio $(M_{\phi^0}/M_{\phi^+})^2$ can be $\sim 7$ (at 90\% C.L.) 
in the most conservative case with $M_{\phi^+} \sim 50$ GeV   
 \cite{BGP,PDG}.  
Therefore  the effective parameter $\eps_{\tau R}$  
of ${\cal L}^\nu_{eff}$ 
can be at most a factor 7 larger than the present experimental constraint  
on the effective $\kappa_{\tau R}$ coupling, $\kappa_{\tau R}\lsim 0.1$ 
(see next Section).

\noindent
$\bullet$  We can exchange   
 some extra singlet fermion  $N$ according to these interactions:
\be{NN}
h \bar{l}_\tau P_R N H   + M_N \bar{N} N + f \bar{N}P_R e S +{\mathrm h.c.} ,
\ee
where $S$ is some charged scalar $SU(2)_W$ singlet,  $M_N$ 
is the Dirac mass of the fermion $N$ and $h,f$ are dimensionless 
coupling constants. 
In this case, after decoupling the fields $N$ and $S$, we obtain 
the  $d=8$ operator $(l^\dagger_\tau H)(H^\dagger l_\tau) (e^\dagger e)$, 
which is a linear combination of the two ones shown in (\ref{Rinv}). 
This induces the interaction ${\cal L}^\nu_{eff}$ with 
$\Lambda \sim \sqrt{M_N M_S}$. 
However, the decoupling of $N$ also generates the $d=6$ operator  
$(h/M_N)^2(l^\dagger_\tau H)
\gamma^\mu \partial_\mu (H^\dagger l_\tau)$, which induces a correction 
$(h v/M_N)^2 \nu_\tau^\dagger 
\gamma^\mu \partial_\mu  \nu_\tau$ to the neutrino kinetic term. 
The consequent  redefinition of the $\nu_\tau$ field gives rise to 
extra (charged current as well as neutral current) neutrino interactions
of relative strength $\sim (h v/M_N)^2$. As we will see below, 
the strongest bound comes from the $\tau$-decay which now  
imposes  $M_N/h \gsim 10 ~v$. 
Notice that the interactions in (\ref{NN})
cannot  induce Majorana  neutrino mass terms 
and therefore the mass scales $M_S$ and $M_N$  
involved are not forced to  be much larger than 
the electroweak scale. We can even have $M$ not so far 
from the weak scale and  
so $\eps_{\tau R}$ can be $\sim 0.1 \div 1$.

\section{Laboratory bounds on  NS interactions}
The most stringent constraints on the `observable' interactions 
for the first generation come from   
the collisions  $e^+ e^- \rightarrow e^+ e^-$  
at LEP \cite{PDG} which then constrain the interaction in 
${\cal L}^{(1)}_{eff}$ (at 95\% C.L.):
\beqn{eelep}
 && \kappa_{e R}\lsim 0.01 , ~~~~~ \Lambda > 5.3~ {\mathrm TeV} , 
\nonumber \\
&& \kappa_{e L}\lsim 0.03 , ~~~~~ \Lambda > 3.8~ {\mathrm TeV} ,  
\eeqn
where now the cut-off scale is defined, for given parameter 
$\kappa_{e R (L)}$, according to 
$2\sqrt{2}G_F\kappa_{e R(L)}\equiv 4\pi/\Lambda^2$.
In what follows, the bounds on the cut-off scales will be inferred 
in analogous way also for other interactions.\footnote{
Notice that these $\Lambda$ parameters, defined in order to 
conform with the common practice, are different from the scale 
$M$ introduced in the previous section.} 
As for the third generation, analogously, 
from the collisions  $e^+ e^- \rightarrow \tau^+ \tau^-$ at LEP 
one also infers the most severe 
constraints on the couplings $\kappa_{\tau R,L}$ (at 95\% C.L.): 
\beqn{etau} 
&&\kappa_{\tau R}\lsim 0.1 , ~~~~~~ \Lambda > 2 ~{\mathrm TeV} , 
\nonumber \\
&& \kappa_{\tau L}\lsim 0.02 , ~~~~~~ \Lambda > 4 ~{\mathrm TeV} . 
\eeqn
On the other hand,  the strongest 
bound on the third-generation 
is that on the interaction ${\cal L}^{(2)}_{eff}$ (\ref{NS-2}) 
that is  on the parameter $\zeta_{\tau L}$. We obtain 
from the measurement of the ratio 
$\frac{B(\tau \to \nu_\tau e  \bar{\nu}_e)}{
B(\tau \to \nu_\tau \mu  \bar{\nu}_\mu)} = 0.974\pm 0.005$ 
 \cite{PDG}:
\be{tauL}
|\zeta_{\tau L}| \lsim 2.6\cdot 10^{-3} , ~~~~~ \Lambda > 11 ~{\mathrm TeV} . 
\ee

As anticipated in the previous Section, the $\nu_\tau$-interaction in 
$ {\cal L}^\nu_{eff}$ 
is not experimentally  constrained 
and also the $\nu_e$-interaction
is poorly tested by low-energy $\nu e$ scattering experiments.
Therefore we shall first review the bounds imposed by 
$\nu_e e$ and $\bar{\nu}_e e$ scattering experiments.
In  presence of extra interactions, with the effective 
couplings (\ref{NCnutau}), the cross-section of the 
$\nu_e e$ elastic scattering becomes 
(for $m_e \ll E_\nu \ll M_W$): 
\be{sigma} 
\sigma_{\nu_e e} = \frac{2}{\pi} G_F^2 m_e E_\nu 
\left[ (\tilde{g}_{eL} + 1)^2 + \frac13 \tilde{g}_{eR}^2 \right] .
\ee 
\begin{figure}[h]
\vskip 0.23cm
\hglue -0.82cm
%\centerline{\protect\hbox{
\epsfig{file=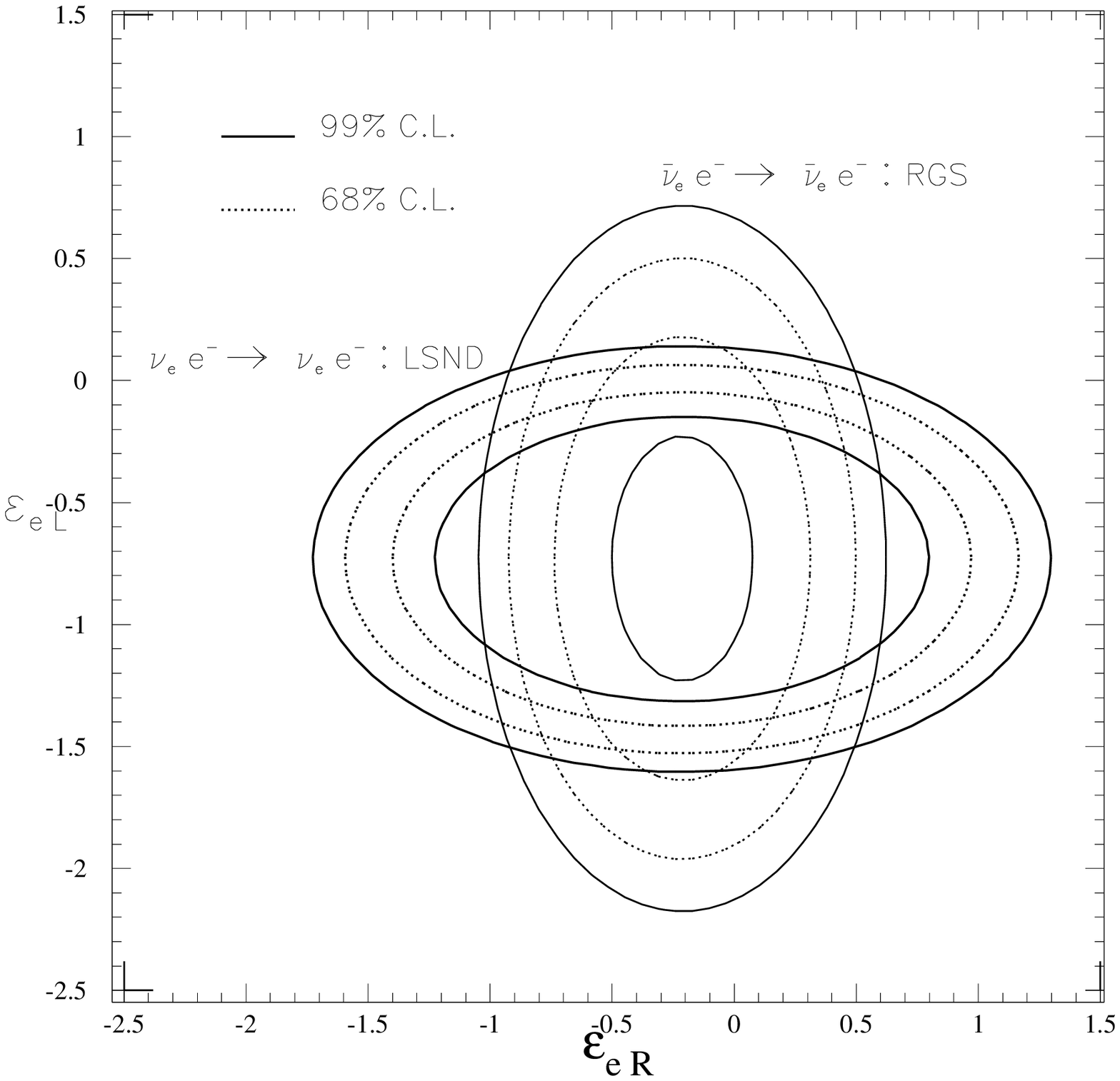,height=7.5cm,width= 8.4cm}
%\framebox[55mm]{\rule[-21mm]{0mm}{43mm}}
\vglue -7.4cm 
\hglue 7.1cm 
\epsfig{file=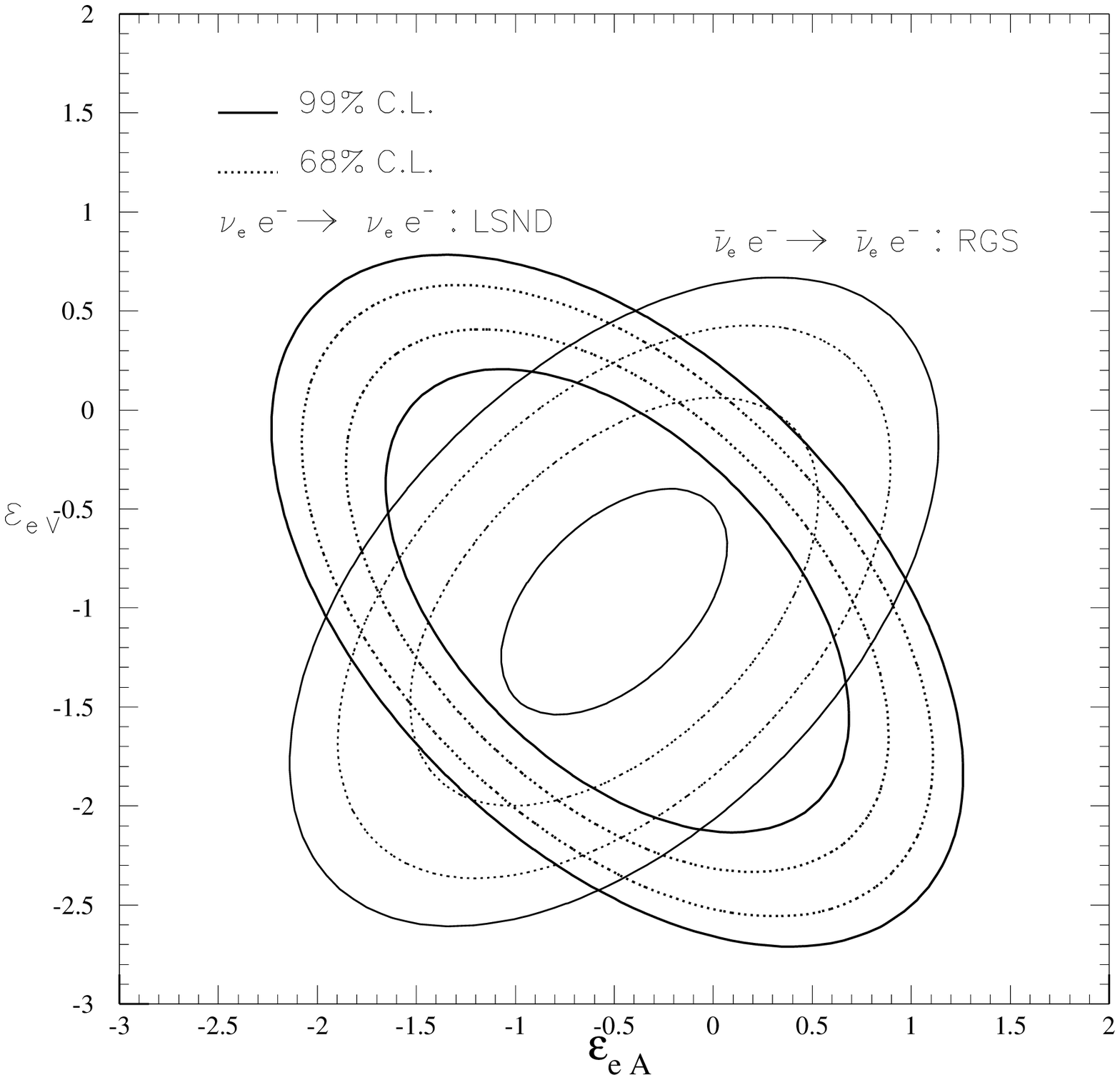,height=7.5cm,width= 8.4cm}
%,angle=270}
\vskip -0.3cm
\caption{\small 
Sensitivity contours to non-standard  $\nu_e$ interactions in the plane 
($\eps_{e R},\eps_{e L}$) (left panel) or 
($\eps_{e A},\eps_{e V}$) (right panel) from the LSND measurement of the  
$\nu_e~ e^- \to \nu_e~ e^-$ cross section and the RGS 
measurement of the  
$\bar{\nu}_e~ e^- \to \bar{\nu}_e~ e^-$ cross section. 
The present 1-$\sigma$ accuracy is 
of 15\% from LSND experiment and 29\% from RGS experiment 
(see more details in the text).
}
%\vskip -0.5cm
\label{fg1}
\end{figure}
The most accurate measurement of this cross section has been 
performed by the LSND collaboration \cite{nue-e}:
\be{ls}
\sigma_{\nu_e e}^{\rm exp} = 
\left(10.1\pm 1.5\right)\cdot E_{\nu}
[{\mathrm MeV}]
\cdot 10^{-45}~{\mathrm cm}^2 .
\ee
In Fig.~\ref{fg1} we draw the iso-contour of sensitivity to the 
NS couplings ($\eps_{eR},\eps_{e L}$) (left panel) or ($\eps_{eA},
\eps_{e V}$) (right panel). We see that a  rather vast parameter region 
is allowed,  where the strength of  the extra interactions 
is larger  than $G_F$.
Notice, that the allowed regions are elliptical annuli 
and it would be misleading to ignore correlations 
between $\eps_{e R}$ and $\eps_{e L}$. 
For example, if we  
 focus on the neighborhood of the 
Standard Model point, $(\eps_{e R},\eps_{e L})=(0,0)$,  
%and visualize the  individual bounds on these parameters.  
we obtain at the $68\%$ or $99\%$ C.L. the following ranges:
% deviations from (\ref{ls}):  
\beqn{limits} 
&& 
{\rm for} ~ \eps_{eR}=0:  ~~~ 
-0.04 \leq \eps_{e L} \leq 0.08 ~~ (68\%) ,  
~-0.15 \leq \eps_{e L} \leq 0.17 ~~ (99\%)  
\nonumber \\ 
&& 
{\rm for} ~ \eps_{eL}=0:  ~~~ 
-0.87 \leq \eps_{e R} \leq 0.41 ~~ (68\%) , 
~-1.16 \leq \eps_{e R} \leq 0.70 ~~ (99\%)  .
\eeqn 
We then observe that the limits on $\eps_{eL}$ are relatively strong,  
while those on  $\eps_{eR}$ are loose enough, $\eps_{eR} \sim 1$ is allowed. 
%% Reactor Part

Now we make use of 
the data from the scattering of 
reactor $\bar{\nu}_e$ off electrons  \cite{reines}.
The theoretical cross section $\sigma_{\bar{\nu}_e e}$ 
is obtained just by exchanging 
$\tilde{g}_{eR} \leftrightarrow \tilde{g}_{eL} + 1$
in eq.~(\ref{sigma}).
The earliest experiment cited in \cite{reines}, 
that by Reines, Gurr and Sobel 
(RGS),  still remains the most accurate\footnote{
Better accuracy is expected in one year or so 
from MUNU experiment \cite{CB}.}. 
The results by the RGS experiment 
have been later reanalyzed
in \cite{Vogel} where a better-understood 
reactor $\bar{\nu}_e$ spectrum is used to derive the cross section.
They found that   
the RGS cross sections   \cite{reines} convoluted with 
the improved $\bar{\nu}_e$ spectrum were 
$1.35\pm 0.4$ and $2.0\pm 0.5$ times the Standard model
prediction, for  recoil electron energy in the intervals 
$1.5 - 3$ MeV  and $3 - 4.5$ MeV, respectively.
 By combining the two data, we may conclude 
that the total cross section (i.e. for recoil electron energy in the 
range $1.5 - 4.5$ MeV) is
$\sigma^{\rm exp}_{\bar{\nu}_e e} =
(1.7\pm 0.5)\sigma^{\rm SM}_{\bar{\nu}_e e}$.
The corresponding limits on $\eps_{e R,L}$ (or $\eps_{e A,V}$)   
are shown in Fig.~\ref{fg1}. 
We see that the bounds on $\eps_{eL}$ are much worse than 
those obtained from $\nu_e e$ scattering, however the limits
on $\eps_{eR}$ are somewhat more stringent. 
Namely, in 
the neighborhood of the point $(\eps_{e R},\eps_{e L})=(0,0)$
we get:\footnote{By combining 
those two data, we implicitly exclude the possibility 
that the discrepancy of the  measurements in the two 
distinct energy ranges be ascribed to   
 strong spectral deformations. In fact, 
this assumption is reasonable in view of 
the LSND limits that prevent $\tilde{g}_{eL}$ to  significantly deviate from  
the SM expectation. Therefore the limits we obtain are certainly reliable 
along the direction $\eps_{e L}=0$.}   
\be{antinue}
{\rm for} ~ \eps_{eL}=0:  ~~~ 
0.08 \leq \eps_{e R} \leq 0.34 ~~ (68\%) , 
~-0.95 \leq \eps_{e R} \leq 0.50 ~~ (99\%) . 
\ee
However, stronger limits on the  parameter $\eps_{e R}$ 
can be obtained by experiments 
sensitive to  the differential cross 
section of the $\nu_e e \to \nu_e e$ scattering at lower 
energies, with $E_\nu \sim m_e$.  
As we have discussed in ref. \cite{BRR} 
the energy spectrum of the recoil electron can be measured 
with good precision at the Borexino detector aimed 
to detect the monoenergetic Beryllium neutrinos from the Sun, 
with $E_\nu =0.86$ MeV.  

\section{The relevance of $\sigma(e^+ e^- \to \nu \bar{\nu}\gamma)$}

Now we would like to come to our main point, namely to derive 
laboratory bounds  from the measurements of 
the $e^+ e^- \to \nu \bar{\nu}\gamma$ cross-section. 
This appears of special relevance for $\nu_\tau$  but we shall show that 
it is relevant also for $\nu_e$.
To our knowledge 
the possibility to extract information on neutrino non-standard interactions 
with electrons from this process has never been discussed in the past. 

In the SM, the reaction $e^+ e^- \to \nu \bar{\nu}\gamma$ 
proceeds via $Z$-boson exchange ($s$-channel) 
and $W$-boson exchange  ($t$-channel). The $W$ contribution is involved 
only for the emission of the pair $\nu_e\bar{\nu}_e$. 
If neutrino non-standard interactions with electrons are present,
they contribute coherently to this reaction.
The total cross section can be written as 
$\sigma = \sigma^{SM}+\sigma^{NS}$, where $\sigma^{SM}$ is the
SM cross section and $\sigma^{NS}$
includes both the pure NS contribution and SM--NS interference.
The constraint $|\sigma - \sigma^{exp}| \leq \delta\sigma^{exp}$, 
where $\sigma^{exp} \pm \delta\sigma^{exp}$ is the experimental result,
can be written in the following form:  
\be{star}
\left| 1 + \frac{\sigma^{NS}}{\sigma^{SM}} 
- \frac{\sigma^{exp}}{\sigma^{SM}} \right|
\leq  \left( \frac{\sigma^{exp}}{\sigma^{SM}} \right)
\left( \frac{\delta\sigma^{exp}}{\sigma^{exp}} \right) \, .
\ee
The ratio $\sigma^{exp}/\sigma^{SM}$ should be evaluated
by combining the latest experimental data with an accurate
computation of the SM cross section (see e.g. \cite{NT,bard}).
On the other hand, as far as NS interactions are concerned, 
for our purposes it is sufficient to compute the ratio 
$\sigma^{NS}/ \sigma^{SM}$ 
using some approximation. In particular, we will work at tree
level and use the `radiator' approximation to describe 
photon emission. Thus we can write \cite{NT}:
\beqn{RA}
\sigma(s)& = & \int {\mathrm d} x  \int {\mathrm d} c_{\gamma}
~H(x,s_\gamma;s)~ \sigma_0(\hat{s}) \, ,
\nonumber \\ 
H(x,s_\gamma;s) &=&
\frac{\alpha}{\pi}\frac{1}{s^2_\gamma}\dfrac{ 1+(1-x)^2}{x}~,
\eeqn
where the cross section $\sigma_0$  
for $e^+ e^- \to \nu \bar{\nu}$, evaluated   
at the energy scale $\hat{s} = (1-x) s$ ($s$ is the centre-of-mass energy), 
is dressed with 
the `radiator' function $H$ expressing the probability to emit 
a photon with an energy fraction $x= 2E_\gamma/\sqrt{s}$ at 
the angle $\theta_\gamma$ ($s_\gamma \equiv \sin\theta_\gamma,\, c_\gamma 
\equiv \cos\theta_\gamma$). Using the SM couplings and the contact
NS interaction (\ref{NS}), we find:
\beqn{crossSM}
\sigma_0^{SM}(s) &= & \dfrac{N_\nu G_F^2}{6\pi} M^4_Z 
( g^2_{R} +
g^2_{L}) \dfrac{s}{\left[ (s-M^2_Z)^2 +(M_Z \Gamma_Z)^2\right]}  
\nonumber  \\
&& +\dfrac{G_F^2}{\pi} M^2_W\left\{ 
 \dfrac{s + 2M^2_W}{2s} 
-\dfrac{M^2_W}{s} \left(\dfrac{s + M^2_W}{s}\right)\log\left(  
\dfrac{s + M^2_W}{M^2_W}\right) 
\right .\nonumber\\
&&- \left.  g_{ L} \frac{M^2_Z (s-M^2_Z)}{(s-M^2_Z)^2 +
(M_Z \Gamma_Z)^2}
\left[ 
\dfrac{(s + M^2_W)^2}{s^2}
\log\left( 
\dfrac{s + M^2_W}{M^2_W}\right) -
\frac{M^2_W}{s} -\frac32\right]
\right\} 
\eeqn
\beqn{crossNS}
\sigma_0^{NS}(s) &= & 
\sum_{\al=e,\mu,\tau} \dfrac{G_F^2}{6\pi}s\left[
(\eps_{\al L}^2 +\eps_{\al R}^2) -
2 (g_{ L}\eps_{\al L} + g_{ R}\eps_{\al R}) 
\dfrac{M^2_Z (s- M^2_Z)}{ (s-M^2_Z)^2 +
(M_Z \Gamma_Z)^2}\right] \nonumber \\
&& +\dfrac{G_F^2}{\pi} \eps_{e L}M^2_W\left[ 
\dfrac{(s + M^2_W)^2}{s^2}
\log\left( 
\dfrac{s + M^2_W}{M^2_W}\right) -
\frac{M^2_W}{s} -\frac32\right] ,
\eeqn
where 
$N_\nu$ is the number of neutrinos (fixed to three),  
$M_{W}, M_Z, \Gamma_Z$ are the $W$-boson mass, the $Z$-boson mass and 
total decay rate, respectively. 
The three lines in eq.~(\ref{crossSM}) originate 
from the square of the $Z$ amplitude, 
the square of the  $W$ amplitude and 
$W-Z$ interference, respectively.   
In eq.~(\ref{crossNS}), the terms quadratic in $\eps_{\al L},\eps_{\al R}$
originate from the square of the NS amplitude, whilst those
linear in $\eps_{\al L},\eps_{\al R}$ originate from the interference
of the NS amplitude with the $Z$ amplitude and (for $\nu = \nu_e$)
with the $W$ amplitude.
We can now plug eqs.~(\ref{crossSM}) and (\ref{crossNS}) into 
eq.~(\ref{RA}), integrate over the photon variables and 
construct the ratio $\sigma^{NS}(s)/ \sigma^{SM}(s)$,
to be compared with the experimental data through eq.~(\ref{star}).
In order to simplify this comparison, we will
tentatively assume that the central value $\sigma^{exp}$ 
coincides with the {\em full} $\sigma^{SM}$. In this
case eq.~(\ref{star}) reads 
\be{sensi}
\left| \frac{\sigma^{NS}(s)}
{ \sigma^{SM}(s)} \right|
\leq  
\dfrac{\delta\sigma^{exp}(s)}{\sigma^{exp}(s)} ,
\ee
which can be translated into bounds for the NS couplings.
Using this procedure, we have obtained bounds for each
specific pair $(\eps_{\al R},\eps_{\al L})$ 
(or equivalently  $(\eps_{\al A}, \eps_{\al V})$), 
setting to zero the other $\eps$ parameters (barring possible  
cancellations among the different couplings).

To obtain the cross sections $\sigma^{NS}(s)$ and $\sigma^{SM}(s)$
through eq.~(\ref{RA}), the integration in the $x$ variable 
has been performed  numerically from $x_{min}=0.1$ to 
$x_{max}=1$. 
We are aware that a better analysis could be done  
by cutting at lower $x_{max}$, below the 
`radiative $Z$-peak return' occurring at $x= 1- {M^2_Z}/{s}$. 
In so doing one could study the interplay between an increase in the  
sensitivity 
to non-standard couplings and an increase in the statistical error 
(due to the loss of events).\footnote{
A better determination of the allowed parameter space would require 
a proper fit to the experimental data. This is beyond the scope of 
our work. Nevertheless  we urge our experimentalist colleagues to accomplish 
this analysis.}

\begin{figure}[ht]
\vskip -0.cm
\hglue -0.82cm
%\centerline{\protect\hbox{
\epsfig{file=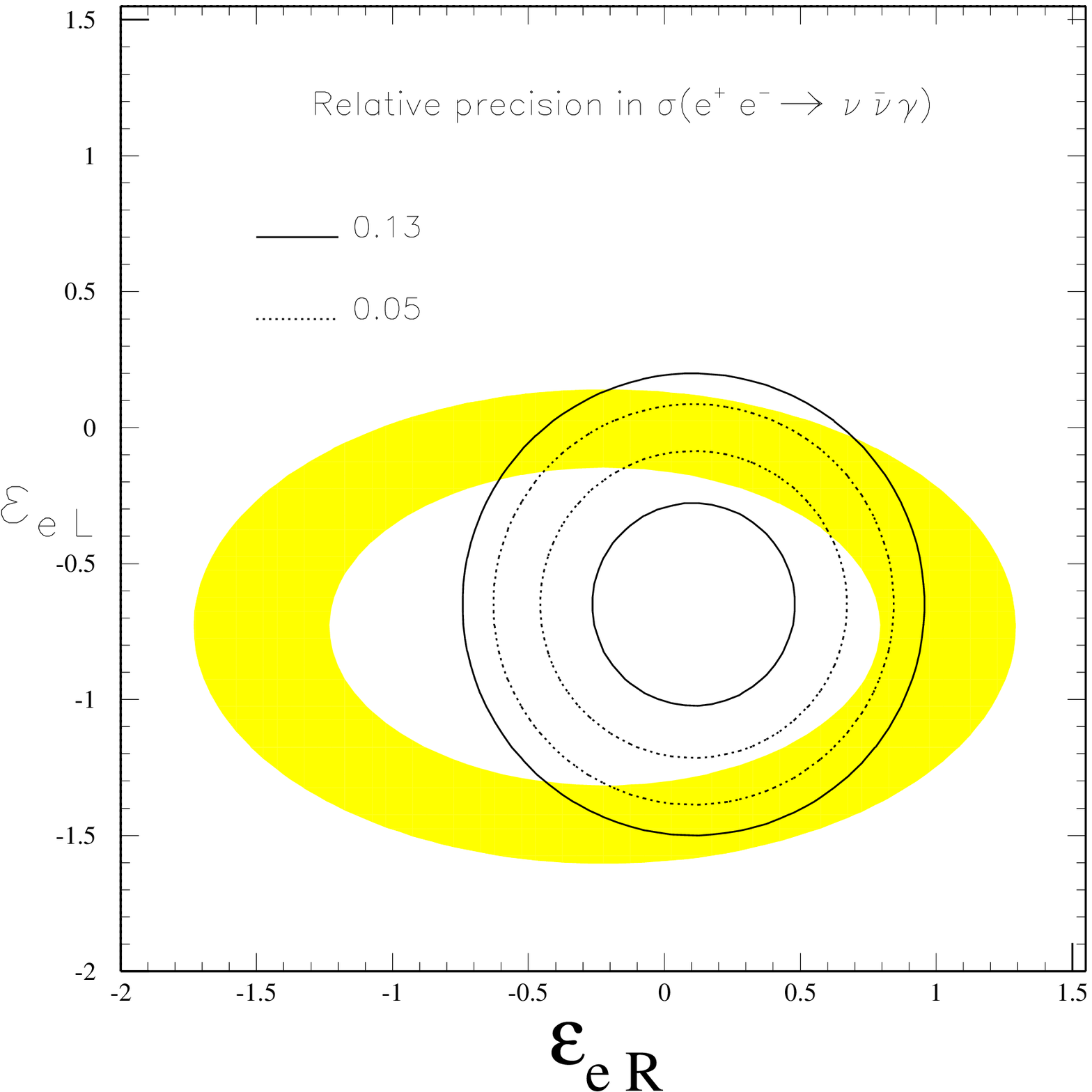,height=7.5cm,width= 8.4cm}
%,angle=270}}}
%\framebox[55mm]{\rule[-21mm]{0mm}{43mm}}
\vglue -7.43cm 
\hglue 7.1cm 
\epsfig{file=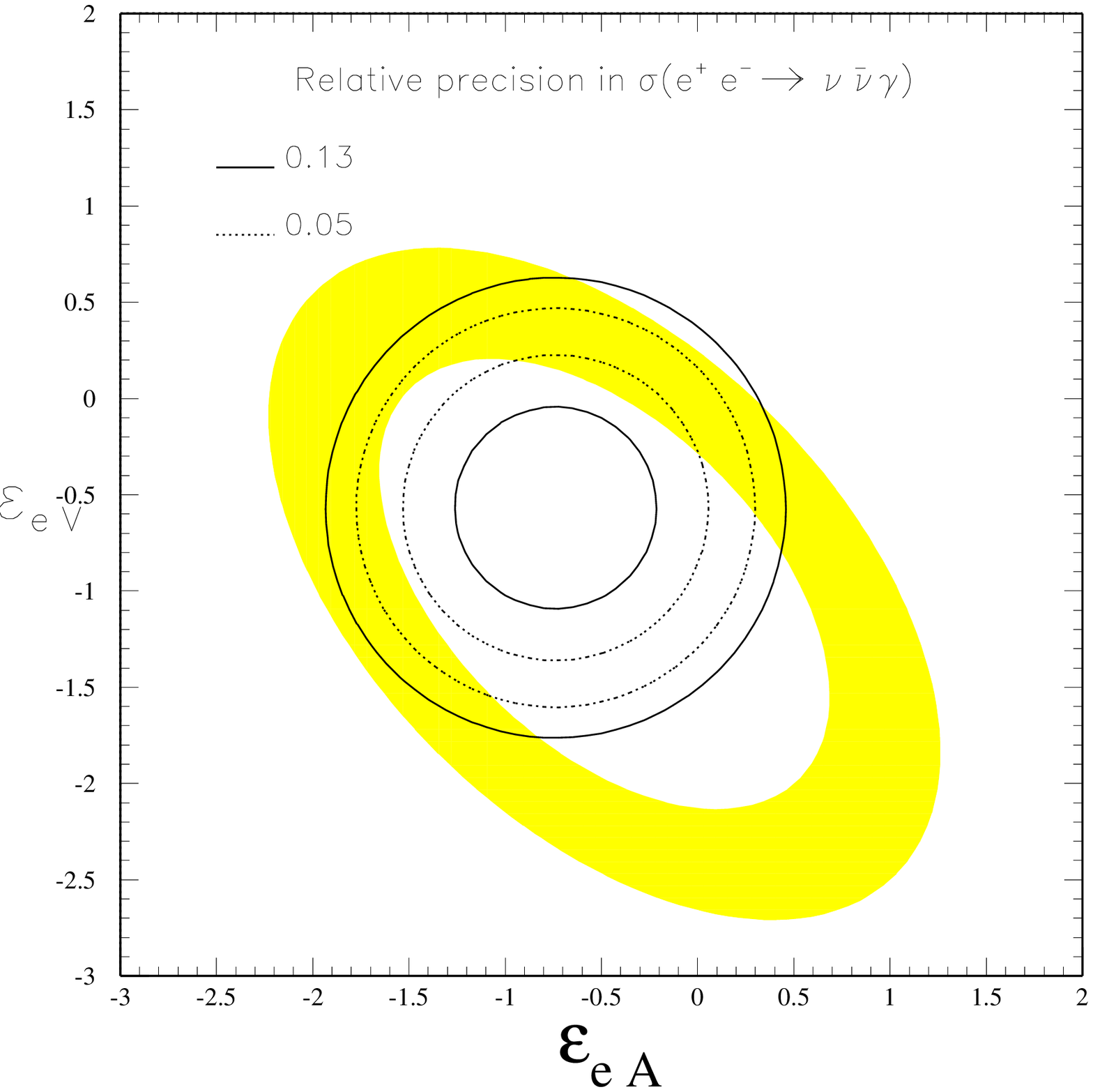,height=7.5cm,width= 8.4cm}
\vskip -0.3cm
\caption{\small 
Sensitivity contours to neutrino non-standard interactions in the plane 
($\eps_{e R},\eps_{e L}$) (left panel) or 
($\eps_{e A},\eps_{e V}$) (right panel) from the 
reaction $e^+e^-\to \nu\bar{\nu}\gamma$ for centre-of-mass 
energy $\sqrt{s}=207$ GeV. The dotted (solid) contours delimit the parameter 
space 
 allowed by LEP at 68\% (99\%) C.L.. For comparison
the 99\% C.L. parameter space allowed by LSND is also depicted 
(shaded annulus). 
}
%\vskip -0.5cm
\label{lep_nue}
\end{figure}  
Our `sensitivity' analysis is  presented in Fig.~\ref{lep_nue} 
and Fig.~\ref{lep}, for $\nu_e$ and $\nu_\tau$, respectively.
In both cases 
in the plane $(\eps_{\al R},\eps_{\al L})$ (left panel) 
or $(\eps_{\al A},\eps_{\al V})$ (right panel) 
we have drawn the iso-contours of the experimental accuracy   
$\dfrac{\delta\sigma^{exp}}{\sigma^{exp}}$ taking $\sqrt{s} = 207$ GeV. 
Let us first discuss the allowed ranges for $\nu_e$ non-standard 
couplings.
%%%%%%%%%%%%%%%   
From the four LEP experiments, we have estimated (perhaps 
{\it underestimated})  
the present 1-$\sigma$ accuracy at the level of $\sim$ 5\% \cite{LEP2}. 
Then adopting a $\sim$ 13\% accuracy at 99\% C.L., we consider   
as  allowed parameter space   the 
annulus bounded by the solid lines in Fig.~\ref{lep_nue}.
If we  
 focus on the neighborhood of the 
Standard Model point, $(\eps_{e R},\eps_{e L})=(0,0)$, 
we obtain at the $68\%$ or $99\%$ C.L. the following ranges:
\beqn{lep-nue}
&& 
{\rm for} ~ \eps_{eR}=0:  ~~~ 
-0.09 \leq \eps_{e L} \leq 0.08 ~~ (68\%) ,  
~-0.28 \leq \eps_{e L} \leq 0.20 ~~ (99\%)  
\nonumber \\ 
&& 
{\rm for} ~ \eps_{eL}=0:  ~~~ 
-0.25 \leq \eps_{e R} \leq 0.45 ~~ (68\%) , 
~-0.46 \leq \eps_{e R} \leq 0.65 ~~ (99\%)  
\eeqn
In Fig.~\ref{lep_nue} we have also shown together the  
parameter space 
for $(\eps_{e R}, \eps_{e L})$ (and $(\eps_{e A}, \eps_{e V})$) 
allowed  at 99\% C.L. 
by  the LSND  data (coloured annulus) for the sake of 
comparison. This should help the reader to catch the correlated 
regions allowed by the two experiments and he/she would notice how the 
combination of the data restricts the allowed regions.
For example, for $\eps_{eR}=0$ the allowed range for $\eps_{eL}$ 
is dictated by LSND experiment (see eq.~\ref{limits}), whereas 
 for $\eps_{eL}=0$ the allowed range for $\eps_{eR}$ is 
mostly restricted by LEP and RGS data, e.g. 
$-0.46\leq \eps_{eR}\leq 0.5$ at 99\% C.L.  

\begin{figure}[htb]
\vskip 0cm
\hglue -0.82cm
%\centerline{\protect\hbox{
\epsfig{file=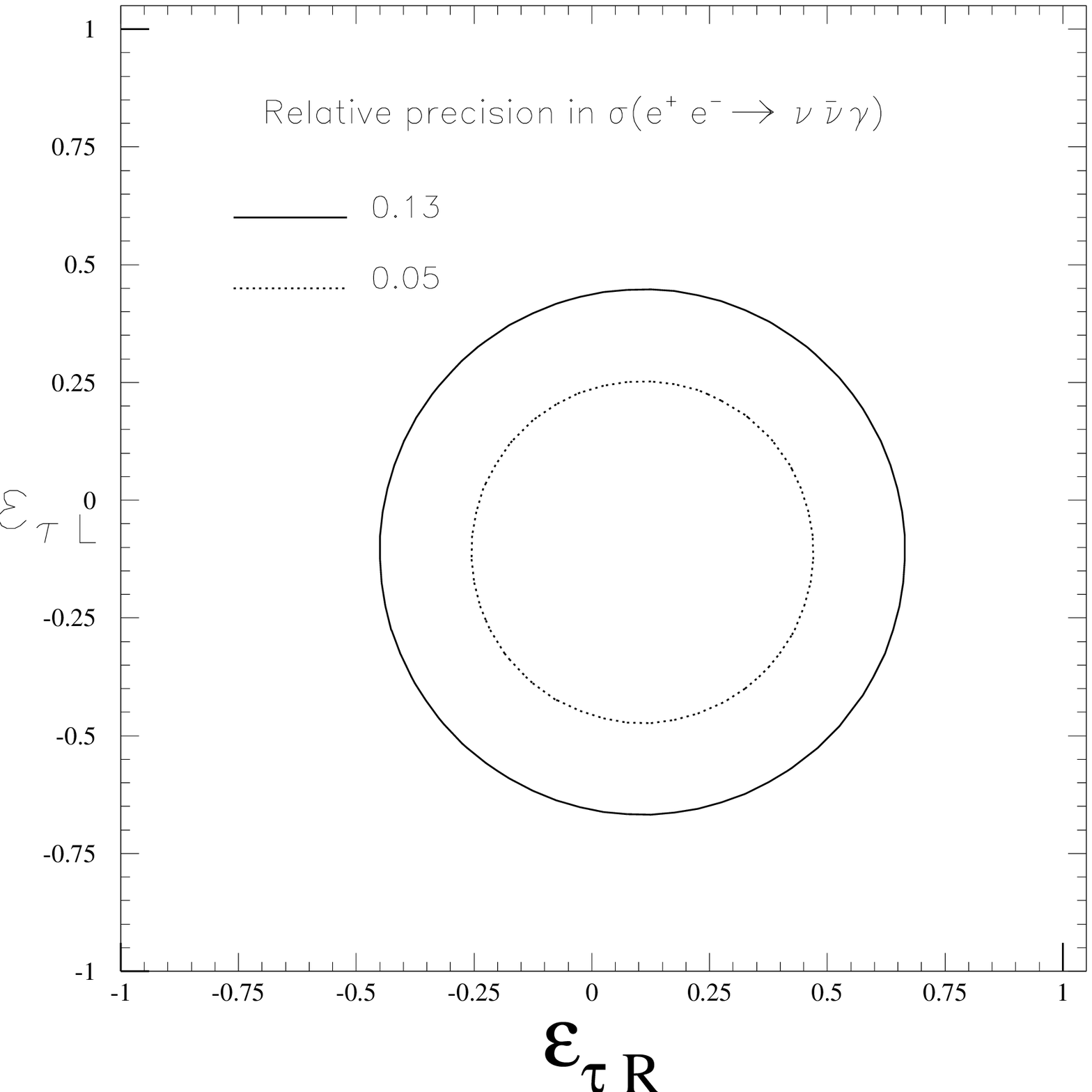,height=7.5cm,width= 8.4cm}
%,angle=270}}}
%\framebox[55mm]{\rule[-21mm]{0mm}{43mm}}
\vglue -7.43cm 
\hglue 7.1cm 
\epsfig{file=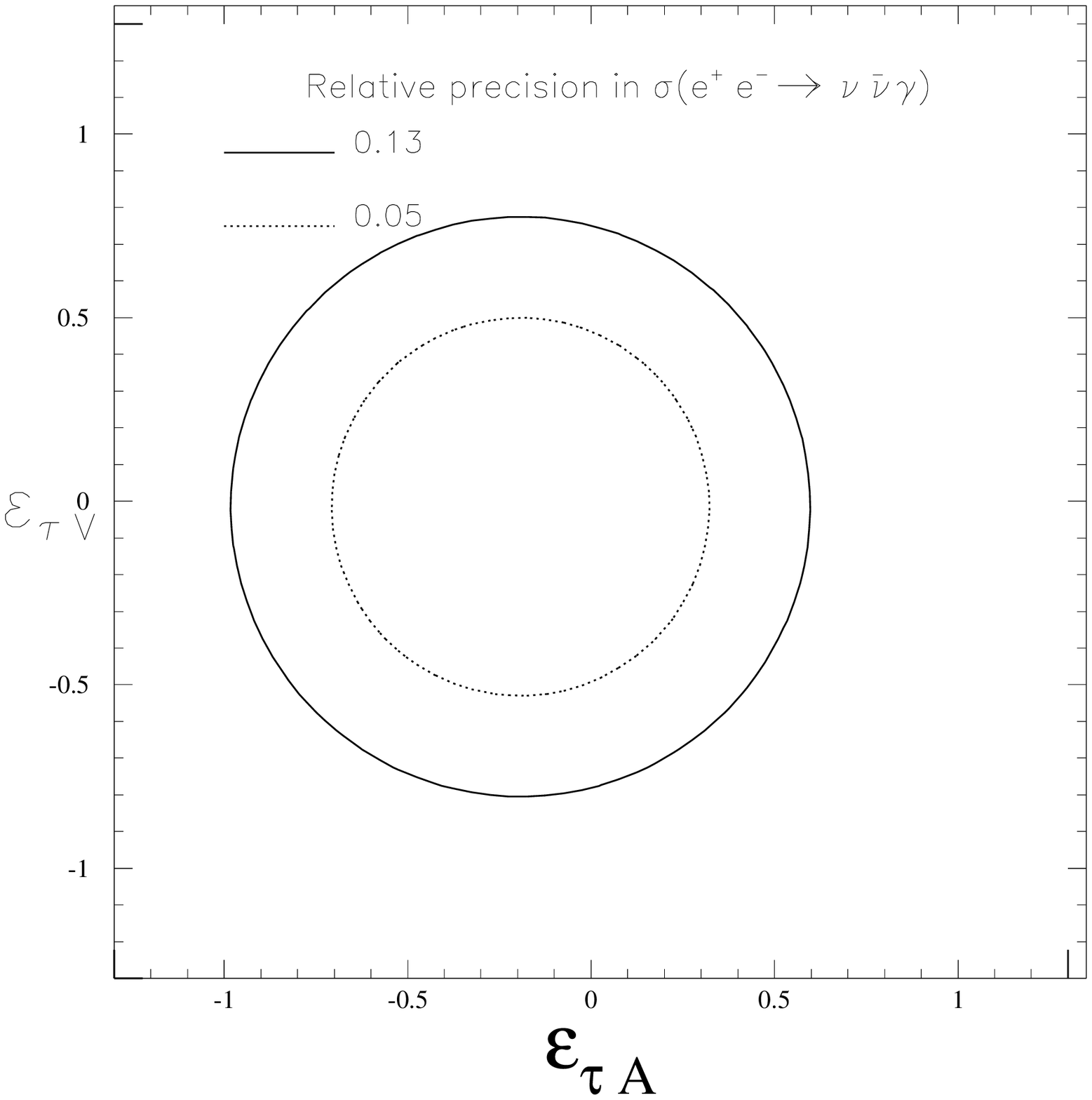,height=7.5cm,width= 8.4cm}
\vskip -0.3cm
\caption{\small 
Sensitivity contours to neutrino NS interactions in the plane 
($\eps_{\tau R},\eps_{\tau L}$) (left panel) or 
($\eps_{\tau A},\eps_{\tau V}$) (right panel) from the 
reaction $e^+e^-\to \nu\bar{\nu}\gamma$ for centre-of-mass 
energy $\sqrt{s}=207$ GeV.
}
%\vskip -0.5cm
\label{lep}
\end{figure}
As for the $\nu_\tau$ non-standard couplings, 
notice from   Fig.~\ref{lep}   that in this 
case the allowed regions become circles
and  so and it  makes more 
sense to ignore the parameter 
correlations in order to find the most conservative bounds\footnote{
We refer as to 
$\nu_\tau$ non-standard interactions, but needless to say that 
the same analysis applies as well  to  $\nu_\mu$ non-standard interactions. 
However, for $\nu_\mu$ these limits are not competitive with 
those from  $\nu_\mu e$ low-energy elastic scattering \cite{charm2}.} 
\be{lep-nutau}
\begin{array}{llll}
 -0.68 \leq \eps_{\tau L} \leq 0.45 , ~~~~~~~~&(\mbox{ any}~ 
\eps_{\tau R}) , \nonumber \\
-0.45 \leq \eps_{\tau R} \leq 0.62 , ~~~~~~~~& (\mbox{any}~ 
\eps_{\tau L}) , \nonumber \\
 -0.82 \leq \eps_{\tau V} \leq 0.70 , ~~~~~~~~&(\mbox {any}~ 
\eps_{\tau A}) 
, \nonumber \\
 -1.0 \leq \eps_{\tau A} \leq 0.55 , ~~~~~~~~&(\mbox{any}~ 
\eps_{\tau V}) .
\end{array}
\ee
%These ranges imply that $\Lambda^+_L > 0.9$ TeV, $\Lambda^-_L > 0.8$ TeV and  
% and $\Lambda^+_R > 0.8$ TeV, $\Lambda^-_R > 0.9$ TeV. 
As is apparent upon comparing the 
sensitivity contours in Fig.~\ref{lep_nue} and Fig.~\ref{lep}, 
for  $\nu_\tau$ the allowed ranges are more symmetric 
around the point ($\eps_{\tau R},\eps_{\tau L})=(0,0)$ 
than those for $\nu_e$ around  ($\eps_{e R},\eps_{e L})=(0,0)$.
This comes from the fact that, at $\sqrt{s}> M_Z$, 
the bounds on $\eps_{\tau R},\eps_{\tau L}$
mainly originate from the terms proportional 
to $\eps^2_{\tau R}, \eps^2_{\tau L}$ in eq.~(\ref{crossNS}), 
which go as $\sim s$. 
In the case of $\nu_e$, the bounds on $\eps_{e R},\eps_{e L}$ 
originate from an interplay between the quadratic terms and the linear
term from $W$ interference. The latter does matter, since its
milder energy dependence ($\sim \log s$) is compensated by a larger
numerical coefficient.

A remark is in order. 
Our expression (\ref{crossNS}) and hence our 
analysis is correct as long as 
 neutrino NS interactions 
 are  {\it point-like} (see eq.~(\ref{NS})). 
However, a modified analysis would be needed in case the effective
NS interactions originate from the exchange of particles 
with masses very close to the experimental energy threshold.
For instance, in the first example discussed  in Sec.~2 (see eq.~(\ref{phi1})),  
the scalar doublet $\phi$ 
is exchanged in the $t$-channel and therefore the corresponding 
diagram for  
$e^+ e^- \to\nu \bar{\nu}$ is similar to that with $W$-exchange. 
In this case for $M_{\phi^+} < 200$ GeV 
the contribution to $\sigma_0^{NS}(s)$ 
quadratic in NS couplings would smoothly pass from a `contact' 
behaviour $\sim s$ for small $s$ to a $1/s$ behaviour 
at higher energies.\footnote{
Notice the different behaviour of the scalar $\phi$-exchange cross 
section  with that from the $W$-exchange: 
the latter exhibits at high energy an energy 
independent behaviour due to the vectorial nature of the vertex. 
Note also that $s$ really means $\hat{s}= (1-x) s$. }   
As a result, we would obtain looser bounds. On the contrary, stronger 
bounds  would be derived 
in case of NS interactions arising from the $s$-channel exchange 
of some extra particle with mass just above 
200~GeV. Indeed, the corresponding contribution would 
start growing faster than $s$ at high energy.
Therefore our results, obtained in the {\it point-like} 
approximation, are representative  from this point of view  
of an intermediate scenario.

Finally, we have to notice that 
the expression in (\ref{crossNS}) is also suitable to 
study the effect of flavour changing neutrino interactions which 
would lead to the process $e^+ e^- \to \nu_{\al}\bar{\nu}_\beta \ga$ 
($\al, \beta =e,\mu,\tau ; ~\al\neq\beta$). 
Analogously to  the flavour conserving couplings discussed until now, 
there are so far no {\it direct} experimental bounds on neutrinos 
flavour-changing couplings with electrons. The existing bounds 
are derived from flavour universality violation and as such they 
apply to the charged operators involving the charged leptons. 
As the process $e^+ e^- \to \nu_{\al}\bar{\nu}_\beta \ga$  
would add incoherently to the 
SM one,  in eq.~(\ref{crossNS}) the interference terms with $Z,W$-diagrams 
should be dropped, and only the quadratic terms in $\epsl^2, \epsr^2$ 
would appear. 
In this case 
the allowed regions are circles centered in (0,0) (at variance with 
those  in Figs.~\ref{lep_nue}, \ref{lep}) and 
the associated most conservative bounds (at 99\% C.L.) are:
\be{lepfc}
\begin{array}{llll}
  |\eps_{\al \beta L }| \leq 0.53 , ~~~~~~~~&(\mbox{any}~ 
\eps_{\al \beta R}) , \nonumber \\
 |\eps_{\al \beta R}| \leq 0.53 , ~~~~~~~~&(\mbox{any}~ 
\eps_{\al \beta L}) , \nonumber \\
  |\eps_{\al \beta V}| \leq 0.75 , ~~~~~~~~&(\mbox{any}~ 
\eps_{\al \beta A}) , \nonumber \\
 |\eps_{\al \beta A}| \leq 0.75 , ~~~~~~~~& (\mbox{any}~ \eps_{\al \beta V}) .
\end{array}
\ee

To conclude this discussion, we add that the astrophysical bounds
from  the stellar evolution imply 
for the strength of neutrino NS interactions with electrons (\ref{NS})    
$\eps_{\al R}, \eps_{\al L} \leq 1$ in the most 
conservative case. 
Neutrino non-standard interactions may affect the primordial 
nucleosynthesis as they would maintain longer these species 
in equilibrium. 
For the sake of completeness we mention that phenomenological bounds 
of NS flavour-changing as well as flavour-diagonal 
neutrino interactions   
have been recently  obtained by atmospheric neutrino data 
fitting \cite{fornengo}. However, we have to stress that these bounds apply  
to the NS vector-coupling $\eps_{\al V}$.   
For the flavour-diagonal coupling  
 with electrons (denoted as 
$\eps'_e$ by the authors \cite{fornengo}) 
we have inferred  that $\eps_{\tau V} \lsim 0.2$ at 99\% C.L..  
Comparable bounds can be deduced from the 
atmospheric neutrino analysis performed 
in \cite{SKS,FLM}. 
The bounds obtained on the amount $\sin^2\xi$ 
 of sterile neutrino  mixed with $\nu_\tau$, namely  $\sin^2\xi < 0.25$ 
\cite{SKS} and 0.8 \cite{FLM},     
can be translated into   $\eps_{\tau V} \lsim 0.125$ and $\lsim 0.4$
at 90\% C.L., respectively. 
However, the analysis performed in \cite{FLM}  
does not rule out the pure sterile 
oscillation which  would imply  $\eps_{\tau V} \lsim 0.5$ at 90\% C.L..

\section{Conclusions}
Extensions of the Standard Model often predict new neutral current 
interactions that can be flavour changing as well as flavour conserving. 
As is well-known, scenarios with neutrino non-standard interactions 
with matter have been invoked to explain the solar and 
atmospheric neutrino anomalies. 
In this paper, we have discussed in detail neutrino NS interactions 
with electrons motivated by the fact that they can be detected 
in Borexino detector through the measurement of the electron 
energy spectrum in the $\nu e$ scattering reaction  \cite{BRR}.
First in Sec. 2.1 we have presented a general operator analysis of such 
non-standard interactions. This  can help to figure out 
the features that the underlying theory has to accomplish 
to fulfill the phenomenological bounds.
Then, the latter have been reviewed in Sec. 3.
The bounds for both the first and third generation are 
tight but they apply to the couplings of the operators involving 
the $SU(2)_W$ related charged-lepton. 
On the other hand, the laboratory limits on $\nu_e$ interactions 
with electrons 
have only been extracted from the measurements of the 
$\nu_e e$ elastic scattering cross section \cite{BPW}. 
We have updated  this analysis using the most recent data from LSND experiment 
and found that the limits are still loose enough (see Fig.~\ref{fg1}). 
In this work we have derived complementary bounds by using  the   
measurements of the 
 $\bar{\nu}_e e$ scattering cross section 
\cite{reines}. 
 As for the third generation, there were no direct bounds on 
 $\nu_\tau$ NS couplings with electrons. 
We have suggested to constrain novel neutrino interactions 
through the reaction $e^+ e^-\to \nu \bar{\nu}\gamma$, measurable at 
$e^+ e^-$ colliders.  
Our results are shown in Figs.~\ref{lep_nue} and \ref{lep}, for 
$\nu_e$ and $\nu_\tau$ case, respectively. All the results 
are obtained considering for given flavour $\al$ ($\al= e, \tau$) 
only the pair $(\eps_{\al R}, \eps_{\al L})$ to be non-zero at time.  
The  accuracy reached by LEP experiments helps to further restrict 
the parameters $\eps_{e R,L}$. By comparing the bounds obtained 
from LSND, RGS and LEP experiment, we can say  that for $\eps_{eL}=0$,       
$|\eps_{e R}|\lsim 0.5$ at 99\% C.L., 
while for $\eps_{e R}=0$,  $-0.15\leq\eps_{e L}\leq 0.17$.
It would be interesting to perform a more refined statistical analysis 
(than that tentatively done by us and displayed in Fig.\ref{lep_nue})  
of these LEP data with those by LSND and possibly by reactor 
$\bar{\nu}_e$, from diffusion on electrons, 
to determine 
accurately the allowed (correlated) parameter space for 
$\eps_{e R}, \eps_{e L}$.    
For $\nu_\tau$ the allowed range looks somehow more symmetric: 
both $|\eps_{\tau R}|$ and  
$|\eps_{\tau L}|$ can be sizeable $\sim 0.5 - 0.7$.
A better accuracy, say $\sim 1\%$,  
in the measurement of the $e^+ e^-\to \nu \bar{\nu}\gamma$ 
cross section \cite{PC}, achievable in the planned Linear Collider \cite{LC}, 
will allow to further shrink the allowed parameter regions. For example, 
in Fig.~\ref{lep_nue} the present annulus allowed by LEP would be replaced 
by a much thiner one (crossing the point (0,0)) and then essentially 
only the upper  overlapping portion  would survive, restricting 
definitely  
$\eps_{e L}$ in the  negative range.\footnote{This holds  if 
no discrepancy with the SM prediction on the cross section is found 
in the experimental data.}  

We have to stress that, at variant with the atmospheric (or solar) neutrino 
phenomenology where only the `vector' parameter $\eps_{\al V}$ can be tested,
 the measurements of $\sigma(e^+ e^- \to \nu \bar{\nu}\gamma)$ 
has allowed to study $\eps_{\al V}, \eps_{\al A}$ 
in a correlated way, and to constrain for the first time 
non-standard couplings of $\nu_\tau$ with electrons.
It would be desirable to exploit further the diagnostic potential 
of  the reaction    
$e^+ e^- \to \nu \bar{\nu}\gamma$ by using  polarized electron - positron 
beams so that to separately disentangle  $\eps_{\al R}$ and $\eps_{\al L}$. 
This also  would be certainly  achieved by  the 
Linear Collider.

\vspace{0.5cm}
{\large \bf Acknowledgments} 
We thank   Andrea Brignole, Carlo Broggini, 
Paolo Checchia and Arcadi Santamaria 
for useful discussions. 
The work of Z.~B. was in part supported by the MURST Research Grant 
"Astroparticle Physics" and that of A.~R.  
by the European Union under the contracts 
HPRN-CT-2000-00149 (Collider Physics) and HPRN-CT-2000-00148 (Across 
the Energy Frontier).


\begin{thebibliography}{9}


\bibitem{FC}
L.~Wolfenstein,
%``Neutrino Oscillations In Matter,''
Phys.\ Rev.\ D {\bf 17} (1978) 2369; \\
J.~W.~F.~Valle,
%``Resonant Oscillations Of Massless Neutrinos In Matter,''
Phys.\ Lett.\ B {\bf 199} (1987) 432; \\
E.~Roulet,
%``Mikheyev-Smirnov-Wolfenstein effect with flavor-changing neutrino  nteractions,''
Phys.\ Rev.\ D {\bf 44} (1991) 935; \\
M.~M.~Guzzo, A.~Masiero and S.~T.~Petcov,
%``On the MSW effect with massless neutrinos and no mixing in the vacuum,''
Phys.\ Lett.\ B {\bf 260} (1991) 154.

\bibitem{AZ1}
Z.~Berezhiani and A.~Rossi,  Phys. \ Rev.\  D {\bf 51} (1995) 5229.
\bibitem{AZ2}
Z.~G.~Berezhiani and A.~Rossi,
%``Matter induced neutrino decay: New candidate for the solution to
% the solar neutrino problem,''
Proceedings of the 5th International  Workshop on `Neutrino Telescopes', p.   
123-135, edt. by M.~ Baldo Ceolin, Venice, Italy, 1993; hep-ph/9306278;  
%Z.~G.~Berezhiani and A.~Rossi,
%``Testing neutrino decay in matter,''
Nucl.\ Phys.\ Proc.\ Suppl.\  {\bf 35} (1994) 469.



\bibitem{analysis}
V.~Barger, R.~J.~Phillips and K.~Whisnant \cite{BPW};\\ 
S.~Bergmann, M.~M.~Guzzo, P.~C.~de Holanda, P.~I.~Krastev and H.~Nunokawa,
%``Status of the solution to the solar neutrino problem based on  non-standard neutrino interactions,''
Phys.\ Rev.\ D {\bf 62} (2000) 073001[hep-ph/0004049] and references 
therein.

\bibitem{nutev}
G.~P.~Zeller {\it et al.}  [NuTeV Collaboration],
%``A Precise Determination of Electroweak Parameters in Neutrino-Nucleon Scattering,''
hep-ex/0110059.

\bibitem{fornengo}
N.~Fornengo {\it et al.}, 
%, M.~Maltoni, R.~T.~Bayo and J.~W.~F.~Valle,
%``Probing neutrino non-standard interactions with atmospheric neutrino  data,''
hep-ph/0108043; \\
P.~Huber and J.~W.~F.~Valle,
%``Non-standard interactions: 
%Atmospheric versus neutrino factory  experiments,''
hep-ph/0108193.



\bibitem{BRR}
Z.~Berezhiani, R.~S.~Raghavan and A.~Rossi, hep-ph/0111138.

\bibitem{BPW}

V.~Barger, R.~J.~Phillips and K.~Whisnant,
%``Solar neutrino solutions with matter enhanced flavor changing neutral current scattering,''
Phys.\ Rev.\ D {\bf 44} (1991) 1629.

\bibitem{sarma}
N.~G.~Deshpande and K.~V.~Sharma, 
Phys.\ Rev.\ D {\bf 43} (1991) 943.\\
%``Limit On Tau-Neutrino Magnetic Moment From Neutrino Counting,'
D.~Fargion, R.~V.~Konoplich and R.~Mignani,
%``Investigation of neutrino properties in the process e+ e- $\to$ neutrino anti-neutrino gamma,''
Phys.\ Rev.\ D {\bf 47} (1993) 751.

\bibitem{DOZ}
A.~D.~Dolgov, L.~B.~Okun and V.~I.~Zakharov,
%``Real And Virtual Photons In Weak Leptonic Processes At High Energies,''
Nucl.\ Phys.\ B {\bf 41} (1972) 197;\\
E.~Ma and J.~Okada,
%``How Many Neutrinos?,''
Phys.\ Rev.\ Lett.\  {\bf 41} (1978) 287
[Erratum-ibid.\  {\bf 41} (1978) 1759];\\
K.~J.~Gaemers, R.~Gastmans and F.~M.~Renard,
%``Neutrino Counting In E+ E- Collisions,''
Phys.\ Rev.\ D {\bf 19} (1979) 1605.

\bibitem{charm2}
P.~Vilain {\it et al.}  [CHARM-II Collaboration],
%``Precision measurement of electroweak parameters from the scattering of muon-neutrinos on electrons,''
Phys.\ Lett.\ B {\bf 335} (1994) 246.



\bibitem{BX}
C. Arpesella {\it et al.} [BOREXINO Collaboration], Proposal of Borexino, 1991 
(unpublished); \\
G.~Ranucci {\it et al.}  [BOREXINO Collaboration],
%``Borexino,''
Nucl.\ Phys.\ Proc.\ Suppl.\  {\bf 91} (2001) 58.

\bibitem{BW}
W.~Buchmuller and D.~Wyler, Nucl.\ Phys.\ B {\bf 268} (1986) 621.
\bibitem{BS}
M.~Bilenkii and A.~Santamaria,
%``Bounding effective operators at the one loop level: The Case of four fermion neutrino interactions,''
Phys.\ Lett.\ B {\bf 336} (1994) 91; 
also in {\it Turin 1999, Neutrino mixing}, p. 50-61,  
%``'Secret' neutrino interactions,''
[hep-ph/9908272].

\bibitem{BGP}
S.~Bergmann, Y.~Grossman and D.~M.~Pierce,
%``Can lepton flavor violating interactions explain the atmospheric  neutrino problem?,''
Phys.\ Rev.\ D {\bf 61} (2000) 053005; \\
S.~Bergmann {\it et al.}
%, M.~M.~Guzzo, P.~C.~de Holanda, P.~I.~Krastev and H.~Nunokawa,
%``Status of the solution to the solar neutrino problem based on  non-standard neutrino interactions,''
Phys.\ Rev.\ D {\bf 62} (2000) 073001.

\bibitem{PDG}
D.~E.~Groom {\it et al.}  [Particle Data Group Collaboration],
%``Review of particle physics,''
Eur.\ Phys.\ J.\ C {\bf 15} (2000) 1.


\bibitem{nue-e}
L.~B.~Auerbach {\it et al.}  [LSND Collaboration],
%``Measurement of electron-neutrino electron elastic scattering,''
Phys.\ Rev.\ D {\bf 63} (2001) 112001.



\bibitem{reines}
F.~Reines, H.~S.~Gurr and H.~W.~Sobel,
%``Detection Of Anti-Electron-Neutrino E Scattering,''
Phys.\ Rev.\ Lett.\  {\bf 37} (1976) 315; \\
G.~S.~Vidyakin {\it et al.},
%``Limitations on the magnetic moment and charge radius of the electron-anti-neutrino,''
JETP Lett.\  {\bf 55} (1992) 206
[Pisma Zh.\ Eksp.\ Teor.\ Fiz.\  {\bf 55} (1992) 212]; \\
A.~I.~Derbin {\it et al.}, 
%A.~V.~Chernyi, L.~A.~Popeko, V.~N.~Muratova, G.~A.~Shishkina and S.~I.~Bakhlanov,
%``Experiment on anti-neutrino scattering by electrons at a reactor of the Rovno nuclear power plant,''
JETP Lett.\  {\bf 57} (1993) 768
[Pisma Zh.\ Eksp.\ Teor.\ Fiz.\  {\bf 57} (1993) 755].

\bibitem{CB}
C. Broggini, private communication.
\bibitem{Vogel}
P.~Vogel and J.~Engel, 
Phys.\ Rev.\ D {\bf 39} (1989) 3378.

\bibitem{NT}O.~Nicrosini and L.~Trentadue,
%``Transverse Degrees Of Freedom In QED Evolution,''
Phys.\ Lett.\ B {\bf 231} (1989) 487;\\
%O.~Nicrosini and L.~Trentadue,
%``Structure Function Approach To The Neutrino Counting Problem,''
Nucl.\ Phys.\ B {\bf 318} (1989) 1.

\bibitem{bard}
D.~Bardin, S.~Jadach, T.~Riemann and Z.~Was,
%``Predictions for anti-nu nu gamma production at LEP,''
hep-ph/0110371.


\bibitem{LEP2}
The most recent literature can be found at:\\ 
http://alephwww.cern.ch/ALPUB/oldconf/summer01/36/photon.ps \\
http://delphiwww.cern.ch/$\sim$pubxx/delsec/conferences/summer01/paper$\underline{}$eps330\underline{}lp142.ps.gz \\
http://l3www.cern.ch/conferences/Budapest2001/papers/note\underline{}2707
/note\underline{}2707.ps.gz \\
http://opal.web.cern.ch/Opal/pubs/physnote/pn470/pn470.ps.gz 
and see also references therein.

\bibitem{SKS}
A.~Habig  [SuperKamiokande Collaboration],
%``Discriminating between nu/mu <--> nu/tau and nu/mu <--> nu/sterile in  atmospheric nu/mu oscillations with the Super-Kamiokande detector,''
arXiv:hep-ex/0106025.
\bibitem{FLM}
G.~L.~Fogli, E.~Lisi and A.~Marrone,
%``Four-neutrino oscillation solutions of the atmospheric neutrino  anomaly,''
Phys.\ Rev.\ D {\bf 63} (2001) 053008;
%[hep-ph/0009299]; 
%.~L.~Fogli, E.~Lisi and A.~Marrone,
%``Super-Kamiokande atmospheric neutrinos: 
%Status of subdominant  oscillations,''
%hep-ph/0105139.
Phys.\ Rev.\ D {\bf 64} (2001) 093005
\bibitem{PC}
P.~Checchia, Proceedings of the Worldwide Study on Physics and Experiments 
with Future Linear $e^+ e^-$ Collider, Vol.~1, p. 376,
Ed. by E.~Fernandez and A.~Pacheco, Universitat Autonoma de 
Barcelona (1999) [hep-ph/9911208].

\bibitem{LC}
E.~Accomando {\it et al.}  [ECFA/DESY LC Physics Working Group Collaboration],
%``Physics with e+ e- linear colliders,''
Phys.\ Rept.\  {\bf 299} (1998) 1.

\end{thebibliography}
\end{document}